\documentclass[letterpaper,10pt]{article}
\usepackage{osameet2}

\usepackage[utf8]{inputenc}

\usepackage{amsmath,amssymb}

\usepackage{tikz}
\usepackage{pgfplots}

\pgfplotsset{compat=1.14}

\usepgfplotslibrary{colorbrewer}
\pgfplotsset{cycle list/Set1-8}
\pgfplotsset{cycle list/Spectral-9}

\usetikzlibrary{spy}

\usetikzlibrary{matrix}
\usetikzlibrary{chains,scopes,shapes}
\usetikzlibrary{fit}

\usepackage[labelsep=period,justification=centering,font=small]{caption}

\usepackage{siunitx}

\usepackage{acronym}

\newacro{AWGN}{additive white Gaussian noise}
\newacro{EDFA}{erbium doped fiber amplifier}
\newacro{EGN}{extended Gaussian noise}
\newacro{FEC}{forward error correction}
\newacro{GN}{Gaussian noise}
\newacro{MB}{Maxwell-Boltzmann}
\newacro{MI}{mutual information}
\newacro{SNR}{signal-to-noise ratio}
\newacro{QAM}{quadrature amplitude modulation}
\newacro{XPM}{cross-phase modulation}
\newacro{WDM}{wavelength division multiplexing}
\usepackage[pdftex,colorlinks=true,bookmarks=false,citecolor=blue,urlcolor=blue]{hyperref} %

\begin{document}

\title{A Simple Nonlinearity-Tailored Probabilistic Shaping Distribution for Square QAM}

\author{Eric~Sillekens$^1$, Daniel~Semrau$^1$, Gabriele~Liga$^1$, Nikita~A.~Shevchenko$^1$, Zhe~Li$^1$, Alex~Alvarado$^2$, Polina~Bayvel$^1$, Robert.~I.~Killey$^1$, and Domaniç~Lavery$^1$} 
\address{1) Optical Networks Group, Dept. of Electronic \& Electrical Engineering, University College London, UK\\%
2) Signal Processing Systems Group, Dept. of Electrical Engineering, Eindhoven University of Technology (TU/e), NL}
\email{e.sillekens@ucl.ac.uk}

\begin{abstract}
A new probabilistic shaping distribution that outperforms Maxwell-Boltzmann is studied for the nonlinear fiber channel. Additional gains of 0.1 bit/symbol MI or 0.2 dB SNR for both DP-256QAM and DP-1024QAM are reported after 200 km nonlinear fiber transmission.
\end{abstract}


\section{Introduction}
Recently, probabilistic shaping has gained research attention in optical communication systems, in part because it can be used as a rate adaptation mechanism for \ac{FEC}. However, as well as being flexible in coding rate, probabilistic shaping offers a throughput gain for the \ac{AWGN} channel. For the \ac{AWGN} channel, the \ac{MB} distribution is the optimal choice for amplitude ring probabilities \cite{kschischang1993optimal}.

The probability mass function of a MB distribution for a random variable $X$ is given by
\begin{align}
	\mathbb{P}\left[X=x_i\right] &= \frac{\exp (-\lambda|x_i|^2)}{\sum_{j=1}^{M}\exp (-\lambda|x_j|^2)},
\end{align}
where $\lambda$ is optimized for the channel and $\{ x_1,x_2,\ldots,x_M \}$ is the set of constellation symbols. In the field of optical fiber communications, this probabilistically shaped distribution was applied to a 64QAM constellation yielding a reach increase of 40\% \cite{buchali2016rate} and to a 1024QAM constellation signal yielding a throughput increase of 13\% \cite{maher2017constellation}.
Additionally, a successful field trial of probabilistic shaping has recently been conducted \cite{cho2017trans}. However, this constellation is optimal for only the \ac{AWGN} channel. For the nonlinear fiber channel, a different distribution can achieve better performance in reach and throughput.

In this work, we include the constellation excess kurtosis induced \ac{SNR} gain; i.e., that the constellation shape can reduce the nonlinear interference, as analytically predicted by \cite{Dar_13,carena2014egn,poggiolini2015simple}. This method yields distributions that introduce less nonlinear interference than a MB distribution and more importantly outperform it in terms of mutual information, as demonstrated in \cite{renner2017experimental}. 
We initially carry out an optimization of the probability distribution taking into account the SNR dependence on the excess kurtosis. We heuristically derive a general expression without loss of performance. We then present simulation results of a single span \SI{200}{km} system (following the work of \cite{renner2017experimental}) at $5\times$\SI{33}{GBd} and with the modulation format varying from dual polarization (DP) 64QAM to 1024QAM. The simulation results have shown a \ac{MI} gain of \SI{0.1}{bit/symbol} and an additional \ac{SNR} gain of \SI{0.2}{dB}.

\section{Mutual Information Optimization Method}

Most analytical models of the nonlinear fiber transmission consider the nonlinear impairments of the channel to behave like \ac{AWGN} \cite{poggiolini2015simple,carena2014egn,Dar_13}. These models have been shown to be accurate in both simulation and experiment \cite{galdino2016experimental}. Similar to \cite{dar2014on}, we assume that the nonlinear interference noise can be written as $\eta_{\text{tot}}P^3 \approx\left(\eta_1+\eta_2\mathfrak{K}\right)P^3$ with real numbers $\eta_1$ and $\eta_2$, the launch power $P$ and the excess kurtosis $\mathfrak{K} \triangleq \frac{ \mathbb{E}\left[|X|^4\right]}{\mathbb{E}\left[|X|^2\right]^2}-2$ of the complex constellation. Straightforward calculations yield the following relationship between the SNR at optimum launch power between an input distribution A and an input distribution B:
\begin{align}
	\frac{\mathrm{SNR}_\mathrm{opt,A}}{\mathrm{SNR}_\mathrm{opt,B}} &= \left(\frac{1+c\mathfrak{K}_\text{B}}{1+c\mathfrak{K}_\text{A}}\right)^{\frac13}, \label{eq:snrinc}
\end{align}
where $c \triangleq \frac{\eta_2}{\eta_1}$ is a measure for the relative impact of the modulation format on the nonlinear interference.
This work uses a Gaussian constellation as a reference distribution as its excess kurtosis is given by $\mathfrak{K}_\text{Gaussian}=0$.

Based on equation~\eqref{eq:snrinc}, the SNR gain as a function of the input distribution can be included in an AWGN channel model. The mutual information is then calculated using Gauss-Hermite quadrature, as shown in \cite[Section III]{Alvarado2011}. A numerical optimization method has been implemented that maximizes the \ac{MI} for a fixed input distribution and its respective SNR by changing the ring probabilities. The trade off between the \ac{SNR} gain from excess kurtosis and the \ac{MI} loss from suboptimal shaping for \ac{AWGN} will yield a net \ac{MI} gain.
Heuristically, we found that the independent ring optimizations did not find any improvements over%
\begin{align}
	\mathbb{P}\left[X=x_i\right] &=  \frac{\exp( -\nu_1|x_i|^2 -\nu_2|x_i|^4)}{\sum_{j=1}^M\exp(-\nu_1|x_j|^2 -\nu_2|x_j|^4)},
\end{align}
where the optimization parameters $\nu_1$ and $\nu_2$ significantly reduce the computational complexity compared to the unconstrained problem.

\begin{figure}[hbtp]
	\centering
	\begin{tikzpicture}[spy using outlines={rectangle, magnification=2,connect spies}]
		\begin{axis}[
		width=\textwidth,
		height=0.5\textwidth,
		xlabel={SNR$_\mathrm{opt,Gaussian}$ [dB]},
		ylabel={MI [bit/4D-sym]},
		xmin=15,xmax=20,
		ymax=14,
		grid=major,
		cycle list name = Set1-8,
		legend cell align=left,
		legend style={font=\footnotesize},
		every axis plot/.append style={line width=1pt,forget plot},
		legend columns=6,
		]
		
        \addlegendimage{Set1-A,mark=square*,only marks,line width=2pt};
        \addlegendentry{64QAM\enspace};
        
        \addlegendimage{Set1-B,mark=square*,only marks,line width=2pt};
        \addlegendentry{256QAM\enspace};
        
        \addlegendimage{Set1-C,mark=square*,only marks,line width=2pt};
        \addlegendentry{1024QAM};
        
        \addlegendimage{line width=2pt};
        \addlegendentry{Uniform};
        
        \addlegendimage{dotted,line width=2pt};
        \addlegendentry{MB};
        
        \addlegendimage{dashed,line width=2pt};
        \addlegendentry{Optimized};

			\addplot[Set1-A] table[x=SNR,y expr=2*\thisrow{uni64QAM},col sep=comma] {data/mi1020.txt};
			\addplot[Set1-A,dotted] table[x=SNR,y expr=2*\thisrow{MB64QAM},col sep=comma] {data/mi1020.txt};
			\addplot[Set1-A,dashed] table[x=SNR,y expr=2*\thisrow{opt64QAM},col sep=comma] {data/mi1020.txt};

			\addplot[Set1-B] table[x=SNR,y expr=2*\thisrow{uni256QAM},col sep=comma] {data/mi1020.txt};
			\addplot[Set1-B,dotted] table[x=SNR,y expr=2*\thisrow{MB256QAM},col sep=comma] {data/mi1020.txt};
			\addplot[Set1-B,dashed] table[x=SNR,y expr=2*\thisrow{opt256QAM},col sep=comma] {data/mi1020.txt};

			\addplot[Set1-C] table[x=SNR,y expr=2*\thisrow{uni1024QAM},col sep=comma] {data/mi1020.txt};
			\addplot[Set1-C,dotted] table[x=SNR,y expr=2*\thisrow{MB1024QAM},col sep=comma] {data/mi1020.txt};
			\addplot[Set1-C,dashed] table[x=SNR,y expr=2*\thisrow{opt1024QAM},col sep=comma] {data/mi1020.txt};

			\draw[Set1-B,line width=1pt]  (axis cs:17.5,2*6.048036812) -- node[pos=0,anchor=base east,inner sep=0,font=\footnotesize] {\SI{\approx0.1}{bit/4D sym}} (axis cs:18,2*6.048036812);
            \draw[Set1-B,line width=1pt] (axis cs:18,2*6.089603425) -- (axis cs:17.5,2*6.089603425) ;
			
			\draw[Set1-C,line width=1pt] (axis cs:19,2*6.055972881) -- node[pos=0,anchor=base west,inner sep=0,font=\footnotesize] {\SI{\approx0.1}{bit/4D sym}} (axis cs:18,2*6.055972881);
            \draw[Set1-C,line width=1pt] (axis cs:18,2*6.104550154) -- (axis cs:19,2*6.104550154);
			
			\draw[Set1-C,dashed,line width=1pt] (axis cs:18,2*6.5) -- (axis cs:18,2*6.104550154);
			\draw[Set1-C,dotted,line width=1pt] (axis cs:18.18,2*6.104550154) -- (axis cs:18.18,2*6.5);
			\draw[Set1-C,<->,yshift=-2pt,line width=1pt] (axis cs:18.18,2*6.5) -- node[midway,anchor=south,font=\footnotesize] {0.2 dB}  (axis cs:18,2*6.5);
            
            \draw[Set1-C,line width=1pt] (axis cs:18.68,2*6.104550154) -- (axis cs:18.68,2*6.5);
			\draw[Set1-C,<->,yshift=-2pt,line width=1pt] (axis cs:18.18,2*6.5) -- node[midway,anchor=south,font=\footnotesize] {0.5 dB}  (axis cs:18.68,2*6.5);

			\coordinate (insetPosition) at (rel axis cs:0.0,0.99);

			\coordinate (spyviewer) at (rel axis cs:0.95,0.05);
           \draw (axis cs:17.75,11.75) -- (axis cs:17.75,12.5) -- (axis cs:18.75,12.5) -- (axis cs:18.75,11.75) -- cycle coordinate[midway] (southbox);
           
		\end{axis}
        \begin{axis}[footnotesize,width=5cm,
        at={(insetPosition)},anchor={outer north west},
        axis background/.style={fill=white},
        grid=major,
        xmin=0,xmax=5,
        ymin=0,ymax=1e-2,
        xlabel={$|x_i|^2$},ylabel={$\mathbb{P}[X=x_i]$},
        every axis plot/.append style={line width=1pt,forget plot},
        ]
        	\addplot[Set1-B,dotted] gnuplot [raw gnuplot,id=pmb256] {plot 'data/pfig.dat' index 0};
        	\addplot[Set1-C,dotted] gnuplot [raw gnuplot,id=pmb1024] {plot 'data/pfig.dat' index 1};
        	\addplot[Set1-B,dashed] gnuplot [raw gnuplot,id=popt256] {plot 'data/pfig.dat' index 2};
        	\addplot[Set1-C,dashed] gnuplot [raw gnuplot,id=popt1024] {plot 'data/pfig.dat' index 3};
        \end{axis}
        
        \begin{axis}[footnotesize,width=6cm,height=3cm,xshift=5mm,yshift=-2mm,
        at={(spyviewer)},anchor={outer south east},
        axis background/.style={fill=white},
        grid=major,
        xmin=17.75,xmax=18.75,xtick distance=0.25,
        ymin=11.75,ymax=12.5,ytick distance=0.25,
        every axis plot/.append style={line width=1pt,forget plot},
        ]
        	
			\addplot[Set1-B] table[x=SNR,y expr=2*\thisrow{uni256QAM},col sep=comma] {data/mi1020.txt};
			\addplot[Set1-B,dotted] table[x=SNR,y expr=2*\thisrow{MB256QAM},col sep=comma] {data/mi1020.txt};
			\addplot[Set1-B,dashed] table[x=SNR,y expr=2*\thisrow{opt256QAM},col sep=comma] {data/mi1020.txt};
			
			\addplot[Set1-C] table[x=SNR,y expr=2*\thisrow{uni1024QAM},col sep=comma] {data/mi1020.txt};
			\addplot[Set1-C,dotted] table[x=SNR,y expr=2*\thisrow{MB1024QAM},col sep=comma] {data/mi1020.txt};
			\addplot[Set1-C,dashed] table[x=SNR,y expr=2*\thisrow{opt1024QAM},col sep=comma] {data/mi1020.txt};
            \coordinate (northax) at (current axis.north);

        \end{axis}
                    \draw (southbox) -- (northax);
        
	\end{tikzpicture}
	\caption{MI as a function of SNR at optimal launch power a for Gaussian-modulated signal. The optimized distribution provides an additional \SI{0.2}{dB} SNR gain on top of \SI{0.5}{dB} SNR gain achieved by the MB distribution. Inset: The probability mass function of the optimized and MB distribution tailored to SNR$_\mathrm{opt,Gaussian}$=\SI{18}{dB}.}
	\label{fig:snr}
    \vspace{-0.5cm}
\end{figure}
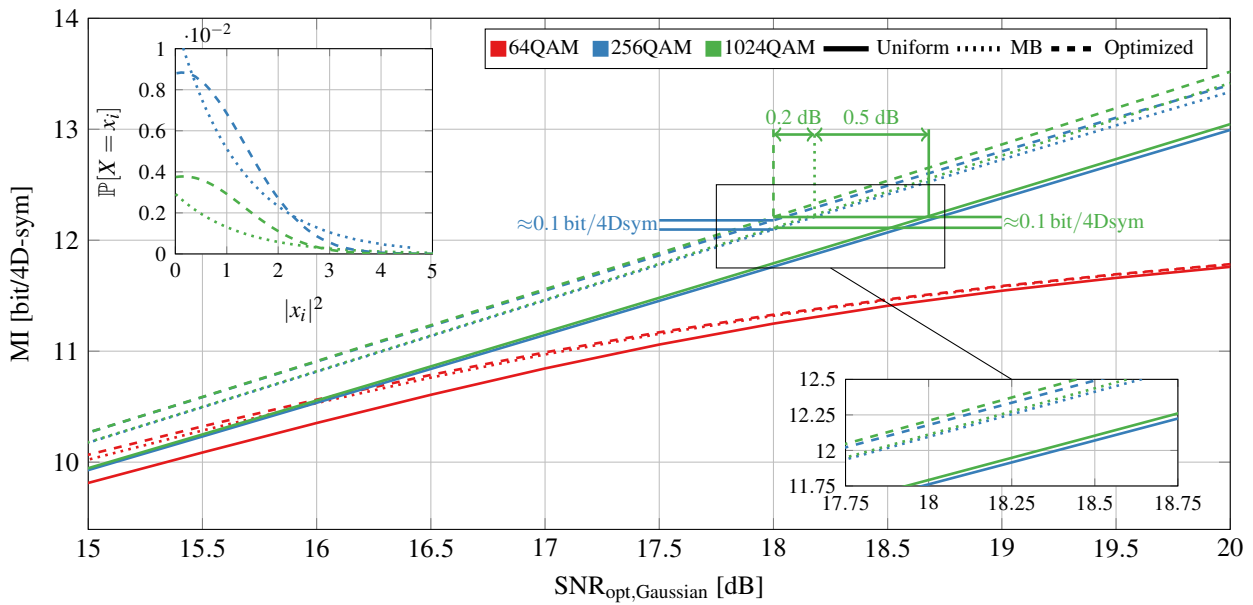

We first evaluated the MI for different square \ac{QAM} modulation formats and optimized the shaped formats for $c=0.69$, a value obtained using equation~\eqref{eq:snrinc} from simulation (see Section \ref{sec:sims}). We assume an \ac{AWGN} channel with the noise variance corrected by the constellation format excess kurtosis. The results of the optimization process are plotted as \ac{MI} versus the optimum \ac{SNR} for a Gaussian-modulated signal as shown in Fig~\ref{fig:snr}. The plotted \ac{MI} values are calculated for an \ac{AWGN} with the \ac{SNR} deviation according to equation~\eqref{eq:snrinc}. The results for DP 64QAM, 256QAM and 1024QAM are shown for uniform distribution, the \ac{MB} distribution and the proposed optimized distribution. For both the \ac{MB} and the proposed optimized distributions, the ring powers are numerically optimized for each value of Gaussian-modulated SNR.

For all modulation formats, the proposed optimized distribution outperforms the \ac{MB} and the uniform distributions for all \ac{SNR} values. For an optimal Gaussian-modulated \ac{SNR} of \SI{18}{dB}, the 64QAM results are already converged. On the other hand, with both DP 256QAM and the 1024QAM, the optimized distribution outperforms the \ac{MB} distribution by approximately \SI{0.1}{bit/symbol} and \SI{0.2}{dB} \ac{SNR}.

\section{Numerical Simulation Demonstration} \label{sec:sims}

We simulated a single span transmission link based on \SI{200}{km} ultra-low-loss single-mode fiber with an attenuation of \SI{0.165}{dB/km}, a dispersion coefficient of \SI{16.3}{ps/nm/km} and a nonlinear parameter of \SI{1.2}{/W/km}, followed by an \acl{EDFA} with a noise figure of \SI{5}{dB}. The transmitter generates a dual-polarization \SI{33}{GHz}-spaced $5\times$\SI{33}{GBd} \ac{WDM} signal, yielding $5\times$\SI{400}{Gbit/s}. The optimum transmission performance is achieved by sweeping the optical launch power per channel. For Gaussian modulation the simulated transmission system achieves an \ac{SNR} of \SI{18}{dB} at optimal launch power.

The simulation results are shown in Fig.~\ref{fig:sims}. The predicted gain from Fig.~\ref{fig:snr} matches well with the results for both 256QAM and 1024QAM. In Fig.~\ref{fig:deltami} it can be observed that the MI gains based on relation \eqref{eq:snrinc} are similar to the results obtained by numerical simulation. This gain is expected to hold for a larger range in SNR.

\begin{figure}[bthp]
	\centering
	\begin{minipage}{.48\linewidth}
	\begin{tikzpicture}
		\begin{axis}[
        	width=\linewidth,
			grid=major,
			xlabel={Launch power per channel [dBm]},
			ylabel={MI [bit/4D-sym]},
			ylabel near ticks,
			xmin=1.5,xmax=7.5,
            ymin=10.75,ymax=12.5,
			legend pos=south east,
			legend cell align=left,
			legend style={font=\footnotesize},
			every axis plot/.append style={line width=1pt},
		]
			\addplot[Set1-B] table[x=power,y expr=2*\thisrow{uni256_MI}, col sep=comma] {data/sim18dB_33GBd.txt};
			\addlegendentry{256QAM Uniform};
			\addplot[Set1-B,dotted] table[x=power,y expr=2*\thisrow{mb256_MI}, col sep=comma] {data/sim18dB_33GBd.txt};
			\addlegendentry{256QAM MB};
			\addplot[Set1-B,dashed] table[x=power,y expr=2*\thisrow{opt256_MI}, col sep=comma] {data/sim18dB_33GBd.txt};
			\addlegendentry{256QAM Optimized};
			
			\addplot[Set1-C] table[x=power,y expr=2*\thisrow{uni1024_MI}, col sep=comma] {data/sim18dB_33GBd.txt};
			\addlegendentry{1024QAM Uniform};
			\addplot[Set1-C,dotted] table[x=power,y expr=2*\thisrow{mb1024_MI}, col sep=comma] {data/sim18dB_33GBd.txt};
			\addlegendentry{1024QAM MB};
			\addplot[Set1-C,dashed] table[x=power,y expr=2*\thisrow{opt1024_MI}, col sep=comma] {data/sim18dB_33GBd.txt};
			\addlegendentry{1024QAM Optimized};

			\draw[Set1-B,line width=1pt] (axis cs:3.5,2*6.07256557949362) -- (axis cs:4.5,2*6.07256557949362);
			\draw[Set1-B,line width=1pt] (axis cs:3.5,2*6.10934048779072) -- (axis cs:4.5,2*6.10934048779072);
			\node[Set1-B,anchor=base east,inner sep=0,font=\footnotesize] at  (axis cs:3.5,2*6.07256557949362) {\SI{\approx0.1}{bit/4Dsym}};
			
			\draw[Set1-C,line width=1pt] (axis cs:5.5,2*6.09210829867075) -- (axis cs:4.5,2*6.09210829867075);
			\draw[Set1-C,line width=1pt] (axis cs:5.5,2*6.13505419861465) -- (axis cs:4.5,2*6.13505419861465);
			\node[Set1-C,anchor=base west,inner sep=0,font=\footnotesize] at  (axis cs:5.5,2*6.09210829867075) {\SI{\approx0.1}{bit/4Dsym}};
*		\end{axis}
	\end{tikzpicture}
	\caption{Simulation results based \SI{200}{km} single span transmission over ultra low loss fiber.} \label{fig:sims}%
	\end{minipage}
    \hfil
\begin{minipage}{.48\linewidth}
    \begin{tikzpicture}
    	\begin{axis}[width=0.99\linewidth,%
        ymin=-0.2,ymax=0.4,
        xmin=0,xmax=20,
        ylabel near ticks,
        xlabel={SNR$_\mathrm{opt,Gaussian}$ [dB]},ylabel={$\Delta$MI [bit/4D-sym]},
        grid=major,
        legend pos=south west,
        legend columns=2,
		legend style={font=\footnotesize},
        legend cell align=left,]
    		\addplot[Set1-A,dotted,line width=1pt] table[x=SNR,y expr=2*\thisrow{mb64}-2*log2(1+10^(\thisrow{SNR}/10))] {data/lowSNRmi.txt};
            \addlegendentry{64QAM MB};
            \addplot[Set1-A,dashed,line width=1pt] table[x=SNR,y expr=2*\thisrow{opt64}-2*log2(1+10^(\thisrow{SNR}/10))] {data/lowSNRmi.txt};
            \addlegendentry{64QAM Optimized};
            
            \addplot[Set1-B,dotted,line width=1pt] table[x=SNR,y expr=2*\thisrow{mb256}-2*log2(1+10^(\thisrow{SNR}/10))] {data/lowSNRmi.txt};
            \addlegendentry{256QAM MB};
            \addplot[Set1-B,dashed,line width=1pt] table[x=SNR,y expr=2*\thisrow{opt256}-2*log2(1+10^(\thisrow{SNR}/10))] {data/lowSNRmi.txt};
            \addlegendentry{256QAM Optimized};
            
            \addplot[Set1-C,dotted,line width=1pt] table[x=SNR,y expr=2*\thisrow{mb1024}-2*log2(1+10^(\thisrow{SNR}/10))] {data/lowSNRmi.txt};
            \addlegendentry{1024QAM MB};
            \addplot[Set1-C,dashed,line width=1pt] table[x=SNR,y expr=2*\thisrow{opt1024}-2*log2(1+10^(\thisrow{SNR}/10))] {data/lowSNRmi.txt};
            \addlegendentry{1024QAM Optimized};
    	\end{axis}
    \end{tikzpicture}
    \caption{MI offset from Gaussian-modulated signals for both MB and optimized distributions.}\label{fig:deltami}
\end{minipage}
\vspace{-0.5cm}
\end{figure}

\section{Conclusion}

We proposed an optimized shaping distribution which provides improved performance over the Maxwell-Boltzmann distribution for the nonlinear channel. We found a simple expression giving an optimized distribution for the nonlinear channel, enabling the optimization of higher order QAM constellations. For both DP-256QAM and DP-1024QAM, the proposed optimized distribution outperforms the Maxwell-Boltzmann distribution by \SI{0.1}{bit/symbol} and provides an additional \SI{0.2}{dB} SNR. This gain is in addition to the \SI{0.5}{dB} SNR gain achieved by shaping using the Maxwell-Boltzmann distribution.

\vspace{0.2cm}
\footnotesize{\textit{%
This work was financially supported in part by the UK Engineering and Physical Sciences Research Council (EPSRC) project UNLOC (EP/J017582/1) and grant (EP/M507970/1) with Neptune Subsea (Xtera), in part by the Netherlands Organisation for Scientific Research (NWO) via the VIDI Grant ICONIC (project number 15685), and in part by the Royal Academy of Engineering under the Research Fellowships Scheme. %
}

\bibliographystyle{osajnl}
\bibliography{shaping}

\end{document}